\documentclass[a4paper,11pt]{article}
\usepackage{jcappub}   

\usepackage{bm}
\usepackage{mathtools}
\usepackage{xcolor}
\usepackage{booktabs}
\usepackage{cleveref}  

\newcommand{\avg}[1]{\langle #1\rangle}
\newcommand{\Dmin}{\Delta_{\min}}
\newcommand{\Dmax}{\Delta_{\max}}
\newcommand{\kmin}{k_{\min}}
\newcommand{\kmax}{k_{\max}}
\newcommand{\Ltot}{L_{\rm tot}}
\newcommand{\Bx}{\avg{B_x^2}}

\title{Domain models of turbulent magnetic fields and the spectral
shape of ensemble averaged axion-photon conversion}

\author[a]{Oleksandr Khasai,}
\author[b,c]{Lidiia Zadorozhna,}
\author[d]{Denys Malyshev}
\author[b]{and Oleg Ruchayskiy}

\affiliation[a]{Bogolyubov Institute for Theoretical Physics of the NASU,
Kyiv, Ukraine}
\affiliation[b]{Niels Bohr Institute, University of Copenhagen,
Copenhagen, Denmark}
\affiliation[c]{Taras Shevchenko National University of Kyiv, Kyiv, Ukraine}
\affiliation[d]{Institut f\"ur Astronomie und Astrophysik T\"ubingen,
Universit\"at T\"ubingen, T\"ubingen, Germany}

\emailAdd{khasai@bitp.kyiv.ua, zadorozhna@nbi.ku.dk, denys.malyshev@astro.uni-tuebingen.de, oleg.ruchayskiy@nbi.ku.dk}

\abstract{Axion-like particles are light pseudoscalars that appear in
many extensions of the Standard Model and are candidates for dark matter.
Through their mixing with photons, they imprint energy dependent
modulations on the spectra of astrophysical sources shining through magnetised media.
When a distant source shines through the turbulent magnetic field of an
intervening galaxy cluster, these modulations are stochastic.
It has recently been shown that averaging
over an ensemble of such sources instead produces a
smooth absorption-like feature.  In this work we derive the analytic form of
this ensemble averaged signal, previously known only numerically.
We further relate the index $a$ of the magnetic field energy spectrum to the
index $\beta$ of the domain size distribution that simulations use to
describe the turbulent field.
For $1<a<2$ the two indices obey the simple analytic relation $\beta=3-a$.
Any distribution with $\beta<1$ instead mimics the spectral index $a=2$,
which the sharp domain walls impose.
For Kolmogorov turbulence, relevant for galaxy clusters, the relation gives $\beta=4/3$.}


\begin{document}

\maketitle
\flushbottom

\section{Introduction}
\label{sec:intro}

The axion was originally proposed to resolve the strong CP problem of quantum
chromodynamics~\cite{Peccei:1977ur,Peccei:1977hh,Weinberg:1977ma,Wilczek:1977pj}.
A broad class of axion-like particles (ALPs) subsequently emerged in
low energy effective theories descending from string compactifications and
other high energy frameworks (see~\cite{Choi:2020rgn,2015ARNPS..65..485G,
2018PrPNP.102...89I} for reviews).  A common feature of axions and ALPs is
their two-photon coupling
\begin{equation}
    \label{eq:alp-photon}
    \mathcal{L}_{\rm int} = g_{a\gamma}\,{\bm E}\cdot{\bm B}\,\phi\,,
\end{equation}
where $\phi$ is the ALP field, ${\bm E}$ and ${\bm B}$ are the electric and
magnetic fields, and $g_{a\gamma}$ is the coupling
strength.  In an external magnetic field this coupling mixes photon and ALP
states~\cite{1988PhRvD..37.1237R,2021RvMP...93a5004S}.
For photons propagating through astrophysical magnetic fields the mixing
produces energy dependent oscillations of the photon survival probability and
hence spectral modulations of background sources.

Galaxy clusters host
$\mu$G-scale magnetic fields coherent over kiloparsec scales along
sightlines of megaparsec length~\cite{Carilli:2001hj,2012A&ARv..20...54F,
2004A&A...424..429M,Bonafede:2010xg}.  X-ray and gamma ray spectra of active
galactic nuclei (AGN) located in or behind clusters have therefore been used
extensively to constrain the ALP--photon
coupling~\cite{2013ApJ...772...44W,Conlon:2017qcw,Marsh:2017yvc,
2018arXiv180504388M,2020ApJ...890...59R,Matthews2022,2022MNRAS.510.1264S}.
Because the cluster field is turbulent, each sightline realises a different
random modulation pattern.  Constraints from individual sources must
therefore be marginalised over many realisations of the magnetic field, which
complicates the searches and weakens the derived limits.
Refs.~\cite{2025NatAs...9.1387M,2025arXiv250922344M} addressed this
obstacle by averaging the conversion probability over many independent
sightlines towards an ensemble of background AGN.  The stochastic
modulations then reduce to a smooth, step-like suppression of the stacked
spectrum, whose shape can be predicted for a given model of the field.

Quantitative predictions for such signals rely on numerical propagation codes,
for example \textsc{ALPro}~\cite{alpro,Matthews2022} or
\textsc{gammaALPs}~\cite{2021arXiv210802061M}, which  represent
the turbulent field as a sequence of \emph{domains} (cells) of uniform field
with random orientations.  The domain sizes $\Delta$ are drawn from a
power law distribution which we parameterise as
\begin{equation}
    p(\Delta)\propto\Delta^{-\beta},\qquad
    \Delta\in[\Dmin,\Dmax] .
    \label{eq:beta_def_intro}
\end{equation}
At the level of the code, $\beta$ is an \emph{ad hoc} input parameter.
Physically, however, the turbulent field is characterised by its
three-dimensional power spectrum, which in the range $\kmin \le k \le \kmax$
takes the power law form
\begin{equation}
    \avg{|B_k|^2}\propto k^{-n},\qquad n = a+2,
    \label{eq:a_def_intro}
\end{equation}
where $B_k$ is the Fourier amplitude of the field at wavenumber $k$.  The
spectral index $a$ of the energy spectrum is measured, e.g., through
Faraday rotation~\cite{2004A&A...424..429M,Bonafede:2010xg,
2012A&A...540A..38V,2017A&A...603A.122G}.  In this notation Kolmogorov
turbulence corresponds to $a=5/3$ ($n=11/3$).

Which value of $\beta$ makes the domain model
reproduce a turbulent field of given $a$?  This question has received
conflicting answers in the literature.  Ref.~\cite{Conlon:2017qcw}
draws domain lengths from $p(\Delta)\propto\Delta^{-1/3}$ and motivates this index by the Kolmogorov power spectrum.  The value follows
from the change of variables $\Delta\sim 1/k$ in the energy spectrum.  Similar
values were adopted in later works, including the ensemble averaged analyses
of~\cite{2025NatAs...9.1387M,2026JCAP...07..066G}.  On the other hand,
ref.~\cite{2020ApJ...890...59R} and the public code
\textsc{ALPro}~\cite{alpro} adopt $p(\Delta)\propto\Delta^{-1.2}$, with the exponent anchored in the observed
three-dimensional rotation measure power spectrum of
Abell~2199~\cite{2012A&A...540A..38V}. 
The two exponents differ by
almost one unit, and neither has been derived from the
statistics of the underlying field.

In this paper we address this discrepancy. 
In the perturbative regime, the ensemble averaged conversion
probability is fully determined by a single function, the line-of-sight
power spectrum of the transverse magnetic field~\cite{Marsh:2021ajy}, or equivalently its Fourier transform, the
two-point correlation function.  We compute this function exactly in
both descriptions and match the results.  Our main results are the
following.
\begin{enumerate}
\item The two-point correlator of the domain model with the size distribution
\eqref{eq:beta_def_intro} is obtained in closed form
(\cref{eq:Kdom}).  Matching it to the correlator of a turbulent field with
spectrum \eqref{eq:a_def_intro} (\cref{sec:matching_slopes}) yields
\begin{equation}
    \boxed{\beta = 3-a} ,
    \label{eq:main_relation_intro}
\end{equation}
valid for $1<a<2$.  For Kolmogorov turbulence $\beta=4/3$.
\item For $\beta<1$ the relation \eqref{eq:main_relation_intro} cannot be
satisfied.  The kink of the domain model correlator at zero separation
generates a universal $k^{-2}$ tail of the line-of-sight power spectrum, and
this tail dominates over the $k^{-(3-\beta)}$ contribution.  A domain model with
$\beta<1$ therefore mimics the effective spectral index $a=2$,
independently of $\beta$.  In particular, the value $\beta=1/3$ adopted
in~\cite{Conlon:2017qcw} falls into this class and does not reproduce the
Kolmogorov scaling (see \cref{sec:literature}).
\item Matching also the amplitudes of the two
power law tails relates the domain size cutoffs to the spectral
cutoffs (\cref{eq:xi_def}).  For $a=5/3$ we find $\kmin \approx 6.33/\Dmax$,
within $1\%$ of the naive identification
$\kmin=2\pi/\Dmax$.  This matching fixes the onset and the inertial range
tail of the averaged signal.  The saturation level at high energies remains
sensitive to the modelling of the largest scales, and we quantify this
ambiguity, of order unity, in \cref{sec:matching_amplitude}.
\item We derive a compact analytic expression, \cref{eq:f_closed}, for the
ensemble averaged conversion probability of the domain model in the limit
$\Dmin/\Dmax\to0$.  The transform for arbitrary $\Dmin/\Dmax$ is given in
\cref{app:transforms}.  We also derive a two power law asymptotic form,
\cref{eq:f_asym}, accurate at the $5$--$10\%$ level near the onset of the
suppression feature and to better than $1\%$ deep in the power law regime.
These expressions provide an analytic counterpart of the empirical
fitting formulae used in~\cite{2025NatAs...9.1387M}.
\end{enumerate}

The paper is organised as follows.  \Cref{sec:master} recalls the
perturbative master formula relating the averaged conversion probability to
the magnetic field correlator.
\Cref{sec:domain} defines the domain model and derives its correlator and
the analytic form of its averaged conversion probability.
\Cref{sec:turbulent} computes the correlator and the line-of-sight spectrum
of a turbulent field with a power law spectrum.  \Cref{sec:matching} matches the two descriptions, derives
\cref{eq:main_relation_intro} and compares it with the prescriptions used in
the literature.
In \cref{sec:interpretation} we interpret the relation physically: a
spectrum dominated by large scales is represented by a size distribution
dominated by small domains.  We also determine where the intuitive argument
behind it, which we call the coverage argument, applies.
\Cref{sec:analytic} gives the energy dependence of the averaged signal and
discusses realistic clusters.  We conclude in \cref{sec:conclusions}.
\Cref{app:qm} derives the master formula and the conditions of its
validity, and \cref{app:transforms} collects the explicit Fourier transforms
and a summary of the notation.

\section{Master formula for the averaged conversion probability}
\label{sec:master}

In this section we recall how the ensemble averaged conversion probability
is expressed through the two-point correlator of the magnetic field.  The
exposition follows refs.~\cite{1988PhRvD..37.1237R,Marsh:2021ajy}.

We consider a photon of energy $\omega$ propagating along the $z$-axis
through a magnetised region of total length $\Ltot$.  We assume that the
field is statistically homogeneous along the path, so that its two-point
function depends only on the separation and its variance is the same
everywhere; the radial profiles of real clusters are discussed at the end of
\cref{sec:analytic}.  The field amplitude itself may fluctuate.  The averaged
conversion probability depends on the distribution of the amplitude only
through its variance, while the full distribution enters the scatter between
sightlines (see below).  We work to leading order in the
ALP--photon coupling $g_{a\gamma}$ and away from any resonances, and we refer
to this as the \emph{perturbative} regime (\cref{app:qm} gives the derivation
and the conditions of applicability).  In this regime the
probability for a photon polarised along $x$ to convert into an ALP of mass
$m_a$ is
\begin{equation}
    P_{\gamma_x\to a}
    = \frac{g_{a\gamma}^2}{4}
      \left|\int_0^{\Ltot} dz\, B_x(z)\, e^{i\Delta_a z}\right|^2,
    \qquad
    \Delta_a = -\frac{m_a^2}{2\omega},
    \label{eq:P_perturbative}
\end{equation}
where $B_x(z)$ is the transverse magnetic field component along the photon
polarisation and $\Delta_a$ is the ALP--photon momentum mismatch.%
\footnote{In a plasma the photon acquires an effective mass equal to the
plasma frequency $\omega_{\rm pl}$, and $\Delta_a$ is replaced by
$\Delta_a+\omega_{\rm pl}^2/(2\omega)$.  For galaxy clusters
$\omega_{\rm pl}\approx1.2\times10^{-12}\,$eV at the typical electron density
$n_e=10^{-3}\,{\rm cm^{-3}}$, so
this correction is negligible for $m_a\gtrsim\text{few}\times10^{-11}\,$eV.}

For a single realisation of a turbulent field, \cref{eq:P_perturbative} is a
rapidly oscillating function of energy.  We are interested in its average over
an ensemble of statistically independent field realisations (in practice, over
many sightlines through the same cluster or a sample of clusters).

We introduce the two-point
correlation function of the transverse field\footnote{The average $\avg{\cdot}$ in
\cref{eq:K_def} is taken over field realisations and, by homogeneity, does
not depend on $z'$.  Equivalently, for a single realisation the same
quantity is the spatial average
$K(z)=\Ltot^{-1}\int_0^{\Ltot}dz'\,B_x(z')B_x(z'+z)$ along the sightline,
which converges to the ensemble average for $\Ltot$ much larger than the
correlation length
(ergodicity).  This is how $K(z)$ is estimated from the simulated
configurations in \cref{fig:correlator}.}
\begin{equation}
    K(z) \equiv \avg{B_x(z')\,B_x(z'+z)} .
    \label{eq:K_def}
\end{equation}
For $\Ltot$ much larger than the correlation length of the field, the
ensemble average of \cref{eq:P_perturbative} becomes
\begin{equation}
    \boxed{
    \avg{P_{\gamma_x\to a}}
    = \frac{g_{a\gamma}^2}{2}\,\Ltot
      \int_0^{\infty} dz\,\cos(\Delta_a z)\, K(z)
    = \frac{g_{a\gamma}^2}{4}\,\Ltot\epsilon_x(\Delta_a)
    } ,
    \label{eq:master}
\end{equation}
where
\begin{equation}
    \epsilon_x(k) = \int_{-\infty}^{\infty} dz\, K(z)\, e^{-ikz}
    \label{eq:eps_def}
\end{equation}
is the \emph{line-of-sight power spectrum} of $B_x$.
In the perturbative regime the averaged conversion probability therefore
depends on the field model only through $K(z)$, or equivalently through
$\epsilon_x(k)$ at $k=\Delta_a$, and two models of the turbulent field give
the same averaged conversion probability if and only if their line-of-sight
correlators coincide.  \Cref{eq:master} agrees with the Fourier formalism
of~\cite{Marsh:2021ajy}.  Ref.~\cite{2017JCAP...01..024K} developed a
statistical treatment of photon--ALP propagation through sequences of
randomly oriented domains, which follows the evolution of both the mean and
the variance of the transfer functions.

The equivalence concerns the ensemble mean only.  The scatter of
$P_{\gamma_x\to a}$ between sightlines involves four-point functions of the
field.  These depend on the distribution of the field amplitude and differ
between models even when their two-point functions coincide; the domain
model in particular is strongly non-Gaussian.  Modelling the
variance of a stacked signal therefore requires more than the relation
between the two models derived here; ref.~\cite{2017JCAP...01..024K} computes
this variance for a domain model with random orientations.  We also note that
replacing the ensemble average by an average over observed sightlines
assumes that the sightlines are statistically independent.  Sightlines
through the same cluster are transversely correlated on scales up to the
largest domain size.  For a cluster at distance $D$, independence
therefore requires angular separations $\gtrsim\Dmax/D$, about one arcminute
for $\Dmax=30\,$kpc at $D=100\,$Mpc.  This condition holds for AGN samples
spread across the cluster face, but it must be checked for close pairs.

Since $\Delta_a$ depends on the photon energy, low energies probe large
$|\Delta_a|$ and thus small spatial scales, while high energies probe large
scales.  For an unpolarised beam the probability is the mean over the two
polarisations.  For the isotropic models considered below
$\avg{B_x^2}=\avg{B_y^2}=\avg{B_\perp^2}/2$, and the unpolarised result equals
\cref{eq:master}.

\section{The domain model: correlator and averaged conversion probability}
\label{sec:domain}

In this section we define the domain model used in numerical propagation
codes (\cref{sec:domain_def}), compute its two-point correlator in closed
form (\cref{sec:domain_correlator}), and test the resulting averaged
conversion probability against Monte Carlo simulations
(\cref{sec:domain_mc}).  In \cref{sec:analytic_dom} we derive this
probability in closed form for a wide range of domain sizes.

\subsection{Definition}
\label{sec:domain_def}

In the domain model the sightline is partitioned into consecutive
segments, called domains, of lengths drawn independently from
the normalised power law distribution
\begin{equation}
    p(\Delta) = C\,\Delta^{-\beta},
    \qquad
    C = \frac{1-\beta}{\Dmax^{1-\beta}-\Dmin^{1-\beta}},
    \qquad \Delta\in[\Dmin,\Dmax],
    \label{eq:p_delta}
\end{equation}
with $\beta\neq1$ (the case $\beta=1$, $p\propto1/\Delta$, can be obtained as
a limit and contains logarithms).  Within each domain the transverse field is
uniform.  The field values in different domains are independent and drawn
from a distribution with zero mean, $\avg{B_x}=0$, and common variance
$\Bx$.  Only this variance enters the ensemble averaged conversion
probability.
The mean domain length is
\begin{equation}
    \avg{\Delta}
    = \frac{1-\beta}{2-\beta}\,
      \frac{\Dmax^{2-\beta}-\Dmin^{2-\beta}}{\Dmax^{1-\beta}-\Dmin^{1-\beta}}
    = \Dmax\,\frac{1-\beta}{2-\beta}\,
      \frac{1-\rho^{2-\beta}}{1-\rho^{1-\beta}},
    \qquad \rho\equiv\frac{\Dmin}{\Dmax}
    \label{eq:mean_delta}
\end{equation}
(we write $\rho$ rather than the customary $\epsilon$ to reserve
$\epsilon_x$ for the power spectrum \eqref{eq:eps_def}).
We note that for $1<\beta<2$ the mean length vanishes in the limit
$\rho\to0$, because the number of domains along the sightline is dominated by ever
smaller ones.  As we show below, the correlator nevertheless remains finite
and non-trivial in this limit.

\subsection{The two-point correlator}
\label{sec:domain_correlator}

\begin{figure}[t]
\centering
\includegraphics[width=0.72\linewidth]{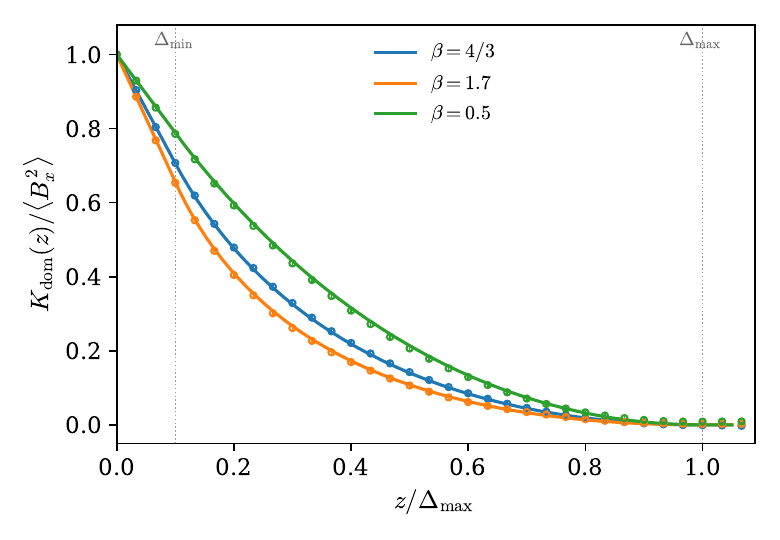}
\caption{Two-point correlator of the domain model for three values of the
size distribution index, $\beta=4/3$ (\emph{blue}), $1.7$ (\emph{orange}) and
$0.5$ (\emph{green}), with $\rho=\Dmin/\Dmax=0.1$.  The dotted vertical lines
mark $\Dmin$ and $\Dmax$.
\emph{Solid lines}: analytic result \eqref{eq:Kdom}; \emph{circles}:
correlators measured from simulated domain configurations ($100$ realisations
of a sightline of length $60\,\Dmax$ each; the field value in each domain is
drawn as $\pm B_{x,\rm rms}$ with equal probability; the mean depends only
on $\Bx$).  Error bars show the standard error of the mean and are mostly
smaller than the symbols.}
\label{fig:correlator}
\end{figure}

The correlator \eqref{eq:K_def} of the domain model can be computed in closed
form.  The key observation is that fields in distinct domains are independent
and have zero mean, so only pairs of points falling in the same domain
contribute:
\begin{equation}
    K_{\rm dom}(z) = \Bx
    \Pi(z),
\end{equation}
where $\Pi(z)$ is the probability that two points separated by $z$ lie in one
domain.  For a stationary sequence of independent lengths, the domain
containing a randomly chosen point follows the \emph{length biased}
distribution $\Delta\,p(\Delta)/\avg{\Delta}$, because longer domains are
proportionally more likely to cover a given point.  Within this domain the
point is uniformly distributed.  The second point then lies in the same
domain with probability $1-z/\Delta$ if $\Delta>z$, and with probability zero
otherwise.  Hence
\begin{equation}
    \Pi(z)
    = \int_{\max(z,\,\Dmin)}^{\Dmax}
      d\Delta\frac{\Delta\,p(\Delta)}{\avg{\Delta}}
      \left(1-\frac{z}{\Delta}\right)
    = \frac{\avg{\left(\Delta-z\right)\theta(\Delta-z)}}{\avg{\Delta}},
    \label{eq:Pi_renewal}
\end{equation}
with $\theta$ the Heaviside step function.  Evaluating the integral for the
distribution \eqref{eq:p_delta} gives
\begin{equation}
    K_{\rm dom}(z) = \Bx\times
    \begin{cases}
    \displaystyle 1 - \frac{z}{\avg{\Delta}},
    & 0 \leq z < \Dmin, \\[10pt]
    \displaystyle
    \frac{\Dmax^{2-\beta} - z^{2-\beta}}{\Dmax^{2-\beta} - \Dmin^{2-\beta}}
    - \frac{z}{\avg{\Delta}}\,
      \frac{\Dmax^{1-\beta} - z^{1-\beta}}{\Dmax^{1-\beta} - \Dmin^{1-\beta}},
    & \Dmin \leq z < \Dmax, \\[10pt]
    0, & z \geq \Dmax .
    \end{cases}
    \label{eq:Kdom}
\end{equation}
The correlator is continuous, decreases monotonically from $K_{\rm dom}(0)=\Bx$,
and vanishes identically beyond the largest domain size, so the domain model
has a strictly finite correlation length.  \Cref{fig:correlator} compares
\cref{eq:Kdom} with correlators measured directly from simulated domain
configurations.  The simulated points follow the analytic curves over the
whole range of separations, for all three values of $\beta$.

We will need two limiting cases of \cref{eq:Kdom}.  First, for a
uniform domain size ($\Dmin=\Dmax=L$, or formally
$\beta\to-\infty$) the correlator reduces to the triangular profile
$K=\Bx(1-z/L)\,\theta(L-z)$, and \cref{eq:master} reproduces the
classic result for domains of equal size,
\begin{equation}
    \avg{P_{\gamma_x\to a}}
    = \frac{g_{a\gamma}^2 \Bx \Ltot}{2}\,L\,
      \frac{1-\cos(\Delta_aL)}{(\Delta_aL)^2}
    = \frac{\Ltot}{L}\,
      \left(\frac{g_{a\gamma} B_{x,\rm rms}\,L}{2}\right)^{2}
      {\rm sinc}^2\left(\frac{\Delta_aL}{2}\right) .
    \label{eq:P0_check}
\end{equation}
This is the number of domains times the single domain conversion probability
familiar from the literature~\cite{2013ApJ...772...44W,Conlon:2017qcw},
$P_0=(\Delta_{a\gamma}L)^2\,{\rm sinc}^2(\Delta_aL/2)$ with
$\Delta_{a\gamma}=g_{a\gamma}B_{x,\rm rms}/2$.

Second, in the limit $\rho\to0$ (a wide range of domain sizes), the region
$z<\Dmin$ shrinks away, and for all $\beta<2$, $\beta\neq1$,
\cref{eq:Kdom} approaches the simple scaling form
\begin{equation}
    K_{\rm dom}(z) \xrightarrow[\rho\to0]{}
    \Bx\left[\,1 - \frac{2-\beta}{1-\beta}\,t
    + \frac{t^{2-\beta}}{1-\beta}\right],
    \qquad t = \frac{z}{\Dmax}\in[0,1],
    \label{eq:Kdom_eps0}
\end{equation}
and zero for $t>1$.  The marginal case $\beta=1$ (log-uniform sizes,
$p\propto1/\Delta$) is the $\beta\to1$ limit of \cref{eq:Kdom_eps0},
\begin{equation}
    K_{\rm dom}^{\beta=1}(z) = \Bx\left(1-t+t\ln t\right),
    \label{eq:Kdom_beta0}
\end{equation}
which we will need below since $a\approx2$ lies within the observationally
allowed range of cluster spectral indices (see,
e.g.,~\cite{2012A&A...540A..38V}).  The matching performed in
\cref{sec:matching} relies on the behaviour of \cref{eq:Kdom_eps0} at small
separations,
\begin{equation}
    1-\frac{K_{\rm dom}(z)}{\Bx}
    = \frac{2-\beta}{1-\beta}\,t - \frac{t^{2-\beta}}{1-\beta} .
    \label{eq:structure_dom}
\end{equation}
The first term is linear in $t$ and analytic, while the second,
$\propto t^{2-\beta}$, is non-analytic.  For $1<\beta<2$ the
non-analytic term dominates as $t\to0$; for $\beta<1$ the linear term
dominates.  As we show in \cref{sec:turbulent,sec:matching}, a turbulent
cascade produces no linear term unless $a=2$, while the $t^{2-\beta}$ term
can be matched to the turbulent structure function.

\subsection{Validation: convergence of the ensemble average}
\label{sec:domain_mc}

\begin{figure}[t]
\centering
\includegraphics[width=0.7\linewidth]{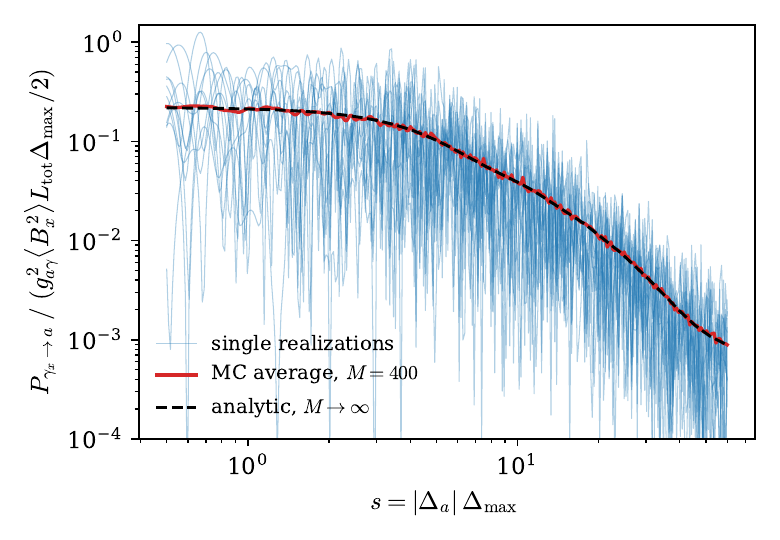}
\caption{Ensemble averaged conversion probability in the domain model
($\beta=1.7$, $\rho=0.1$, $\Ltot=50\,\Dmax$), in units of
$g_{a\gamma}^2\Bx\Ltot\Dmax/2$, as a function of $s=|\Delta_a|\Dmax$.
\emph{Thin blue}: individual realisations; \emph{red}: Monte Carlo average
over $M=400$ realisations ($\pm B_{x,\rm rms}$ domain fields, as in
\cref{fig:correlator}); \emph{dashed black}: analytic result,
\cref{eq:master} with \cref{eq:Kdom}.}
\label{fig:average}
\end{figure}

\Cref{eq:master} together with \cref{eq:Kdom} yields a closed form prediction
for the conversion probability averaged over infinitely many domain
configurations.  \Cref{fig:average} tests this prediction against a direct
Monte~Carlo simulation of the domain model.  Individual realisations
oscillate strongly, while their average converges to the analytic curve.
Here and below we use the
dimensionless variable
\begin{equation}
    s \equiv |\Delta_a|\,\Dmax = \frac{m_a^2 \Dmax}{2\omega} .
    \label{eq:s_def}
\end{equation}
Small $s$ corresponds to high photon energy.  The last equality assumes
$m_a^2\gg\omega_{\rm pl}^2$, see the footnote in \cref{sec:master}.
The Fourier cosine transform of \cref{eq:Kdom} for
arbitrary $\rho$ is elementary but lengthy, and we give it in
\cref{app:transforms}.  Its compact $\rho\to0$ limit is derived in
\cref{sec:analytic_dom}.

\subsection{Analytic form for a wide range of domain sizes}
\label{sec:analytic_dom}

In the limit of a wide range of domain sizes, $\rho=\Dmin/\Dmax\to0$, the
averaged conversion probability of the domain model takes a simple closed
form.  Inserting
\cref{eq:Kdom_eps0} into the master formula \eqref{eq:master} and using the
dimensionless variables $t=z/\Dmax$ and $s=|\Delta_a|\Dmax$
[\cref{eq:s_def}],
\begin{equation}
    \avg{P_{\gamma\to a}}(s)
    = \frac{g_{a\gamma}^2\,\Bx\,\Ltot\,\Dmax}{2} f_\beta(s),
    \qquad
    f_\beta(s) \equiv \int_0^1 dt
    \left(1-\frac{2-\beta}{1-\beta}\,t+\frac{t^{2-\beta}}{1-\beta}\right)
    \cos(st) .
    \label{eq:P_eps0}
\end{equation}
The integral evaluates in closed form:
\begin{equation}
    \boxed{
    f_\beta(s)
    = \frac{2-\beta-s\sin s-(2-\beta)\cos s}{(1-\beta)\,s^{2}}
    +
    \frac{1}{1-\beta}\,
    {\rm Re}\left[
    \bigl(\Gamma(3-\beta)-\Gamma(3-\beta,\,i s)\bigr)\,(i s)^{-(3-\beta)}
    \right]
    } ,
    \label{eq:f_closed}
\end{equation}
where $\Gamma(\alpha,x)$ is the upper incomplete gamma function.  Corrections
to the $\rho\to0$ limit are of relative order $\rho^{2-\beta}$ and are
not negligible for realistic scale ratios.  For $\beta=4/3$ and
$\rho=2/34$ (the value adopted in~\cite{2025NatAs...9.1387M}), where
$\rho^{2/3}\approx0.15$, \cref{eq:f_closed} deviates from
the exact transform by $10$--$15\%$ across the onset region, and the
convergence becomes slower as $\beta\to2$.  For quantitative work at
realistic $\rho$ one should therefore use the exact transform of
\cref{app:transforms} [\cref{eq:I1_app,eq:I2_app}], while
\cref{eq:f_closed,eq:f_asym} describe the shape and the asymptotic
behaviour.

\begin{figure}[t]
\centering
\includegraphics[width=0.7\linewidth]{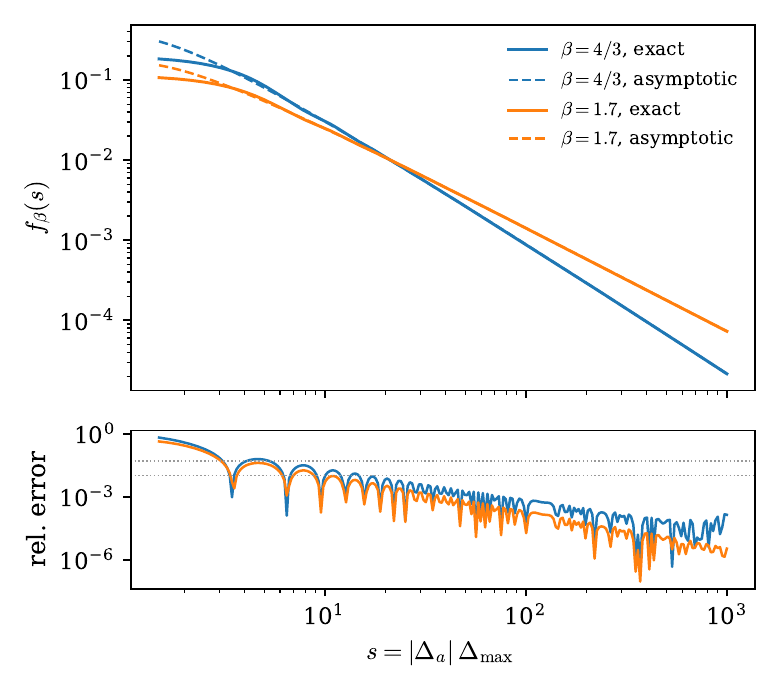}
\caption{\emph{Top}: exact function $f_\beta(s)$ [\cref{eq:f_closed},
\emph{solid lines}] and its two power law asymptotic form \eqref{eq:f_asym}
(\emph{dashed lines}) for $\beta=4/3$ (\emph{blue}) and $\beta=1.7$
(\emph{orange}).  \emph{Bottom}: relative error of the asymptotic formula;
the dotted lines mark $1\%$ and $5\%$.}
\label{fig:asym}
\end{figure}

\Cref{eq:f_closed} involves incomplete gamma functions of imaginary
argument, and its two terms cancel strongly at large $s$.  Expanding
$\Gamma(3-\beta,is)$ for $s\gg1$, we find that all oscillatory terms cancel up
to order $s^{-2}$, which leaves the two power law formula\footnote{The
cancellation follows from the fact that the correlator \eqref{eq:Kdom_eps0}
vanishes together with its first derivative at $t=1$, so that the endpoint
contributions to the transform start at order $s^{-3}$ and are purely
oscillatory.
\Cref{eq:f_asym} therefore contains all non-oscillatory terms of the
large-$s$ expansion.}
\begin{equation}
    \boxed{
    f_\beta(s) \simeq
    -\frac{\Gamma(3-\beta)\sin\frac{\pi\beta}{2}}{1-\beta}
    s^{-(3-\beta)}
    +
    \frac{2-\beta}{1-\beta} s^{-2} ,
    \qquad s\gtrsim \text{few}
    } ,
    \label{eq:f_asym}
\end{equation}
whose first (second) term dominates for $1<\beta<2$ ($\beta<1$); we return
to the two regimes in \cref{sec:matching_slopes}.  In the marginal case $\beta\to1$ the two
terms merge into a logarithm,
\begin{equation}
    f_{1}(s) \simeq \frac{\ln s+\gamma_E}{s^{2}} ,
    \label{eq:f_beta0}
\end{equation}
consistent with the $\beta\to1$ limit of \cref{eq:f_asym}.
\Cref{fig:asym} compares
\cref{eq:f_asym} with the exact transform.  For $\beta=4/3$ the relative error is below $10\%$ for $s\gtrsim3$, below
$5\%$ for $s\gtrsim6$, and below $1\%$ for $s\gtrsim15$.  For $\beta=1.7$ it
is below $5\%$ at $s\gtrsim4$ and below $1\%$ at $s\gtrsim9$.  In the
opposite limit,
\begin{equation}
    f_\beta(0) = \frac{2-\beta}{2(3-\beta)} ,
    \qquad
    f_\beta(s) = f_\beta(0) + \mathcal{O}(s^2),
    \label{eq:f_zero}
\end{equation}
so the averaged probability saturates at high energies.  We express
\cref{eq:f_zero} through the correlation length of the field,
$L_c \equiv (2/K(0))\int_0^\infty K(z)\,dz$, which for the domain model in
the $\rho\to0$ limit is
\begin{equation}
    L_c = \Dmax\,\frac{2-\beta}{3-\beta} .
    \label{eq:Lc}
\end{equation}
This gives $\avg{P}(0)=g_{a\gamma}^2\Bx\Ltot L_c/4$.  At high energies the
beam thus undergoes $\Ltot/L_c$ independent conversions, each with
probability $(g_{a\gamma} B_{x,\rm rms} L_c/2)^2$, in direct analogy with
\cref{eq:P0_check}.  For $\beta=4/3$,
$L_c=(2/5)\,\Dmax$.

\section{Turbulent field with a power law spectrum}
\label{sec:turbulent}

We now compute the same two functions, the line-of-sight correlator and
the power spectrum, for a statistically isotropic, divergence-free turbulent
field with a power law spectrum.  The two-point correlation tensor of such a
field admits the spectral representation
\begin{equation}
    \avg{B_i(\bm{x})\,B_j(\bm{x}+\bm{r})}
    = \int d^3k\,
    \left(\delta_{ij}-\hat k_i \hat k_j\right) \mathcal{M}(k)\,
    e^{i\bm{k}\cdot\bm{r}},
    \qquad
    \mathcal{M}(k)= \frac{3\,(a-1)\,\Bx}{8\pi\left(\kmin^{1-a}-\kmax^{1-a}\right)}
    k^{-(a+2)}
    \label{eq:M_def}
\end{equation}
for $\kmin<k<\kmax$ and $\mathcal{M}=0$ outside.  \Cref{eq:M_def} also fixes our
Fourier convention: the factor $(2\pi)^{-3}$ that accompanies the $k$-space
integral in the standard convention is absorbed into the definition of
$\mathcal{M}(k)$, so that no powers of $2\pi$ appear in \cref{eq:M_def} and
$\Bx=\int d^3k\,(1-\hat k_x^2)\,\mathcal{M}(k)$.  The
transverse projector enforces
$\nabla\cdot\bm B=0$, and the normalisation is fixed by
$K_{\rm turb}(0)=\Bx$, where $K_{\rm turb}$ denotes the correlator
\eqref{eq:K_def} of this field.  The energy spectrum is
$E_B(k)\propto k^2 \mathcal{M}(k)\propto k^{-a}$, so that $a=5/3$ ($n=a+2=11/3$)
corresponds to Kolmogorov turbulence.  We assume $a>1$, so that large scales
dominate the field energy, as observed in
clusters~\cite{2004A&A...424..429M,Bonafede:2010xg}.  We also assume $a<3$, so
that the field remains rough at small separations and its structure function
grows as $z^{a-1}$, see \cref{eq:structure_turb} below.  Any domain
representation requires this roughness: for $a\geq3$ the correlator is
analytic at $z=0$, and the statistics at small separations no longer depend
on the spectral index.

The line-of-sight quantities follow by integrating out the transverse
wavevector components.  For the line-of-sight power spectrum of $B_x$
one finds (see \cref{app:transforms})
\begin{equation}
    \epsilon_x(k_z)
    = 2\pi^2 \int_{\max(|k_z|,\,\kmin)}^{\kmax} \frac{dk}{k}
      \left(k^2+k_z^2\right) \mathcal{M}(k) ,
    \label{eq:eps_turb_int}
\end{equation}
which evaluates to
\begin{equation}
    \epsilon_x(k_z) =
    \frac{3\pi(a-1)\,\Bx}{4\left(\kmin^{1-a}-\kmax^{1-a}\right)}
    \left[
    \frac{k_0^{-a}-\kmax^{-a}}{a}
    + k_z^2\,\frac{k_0^{-(a+2)}-\kmax^{-(a+2)}}{a+2}
    \right],
    \qquad k_0 \equiv \max(|k_z|,\kmin),
    \label{eq:eps_turb}
\end{equation}
for $|k_z|<\kmax$ and zero beyond.  The spectrum has three regimes:
$\epsilon_x$ is flat (up to the $k_z^2$ correction) for $|k_z|<\kmin$;
it vanishes for $|k_z|>\kmax$; and in the inertial range
$\kmin\ll|k_z|\ll\kmax$ it inherits a pure power law,
\begin{equation}
    \epsilon_x(k_z) \simeq
    \frac{3\pi(a-1)(a+1)\,\Bx}{2\,a\,(a+2)\left(\kmin^{1-a}-\kmax^{1-a}\right)}
    |k_z|^{-a}
    \simeq
    \frac{3\pi\,(a^2-1)\,\Bx\,\kmin^{a-1}}{2\,a\,(a+2)}|k_z|^{-a},
    \label{eq:eps_turb_powerlaw}
\end{equation}
where the last expression holds for $(\kmin/\kmax)^{a-1}\ll1$.  The
line-of-sight spectrum falls with the same index $a$ as the energy
spectrum $E_B(k)$, even though the three-dimensional spectrum $\mathcal{M}(k)$ falls as
$k^{-(a+2)}$, as \cref{eq:eps_turb_int} shows; this is a well-known
property of projections of isotropic random fields.  The conversion
probability \eqref{eq:master} involves only the line-of-sight spectrum
$\epsilon_x$.  The domain model, a one-dimensional sequence along the
sightline, has no three-dimensional spectrum, and we match it to the cascade
through the line-of-sight statistics.

\begin{figure}[t]
\centering
\includegraphics[width=\linewidth]{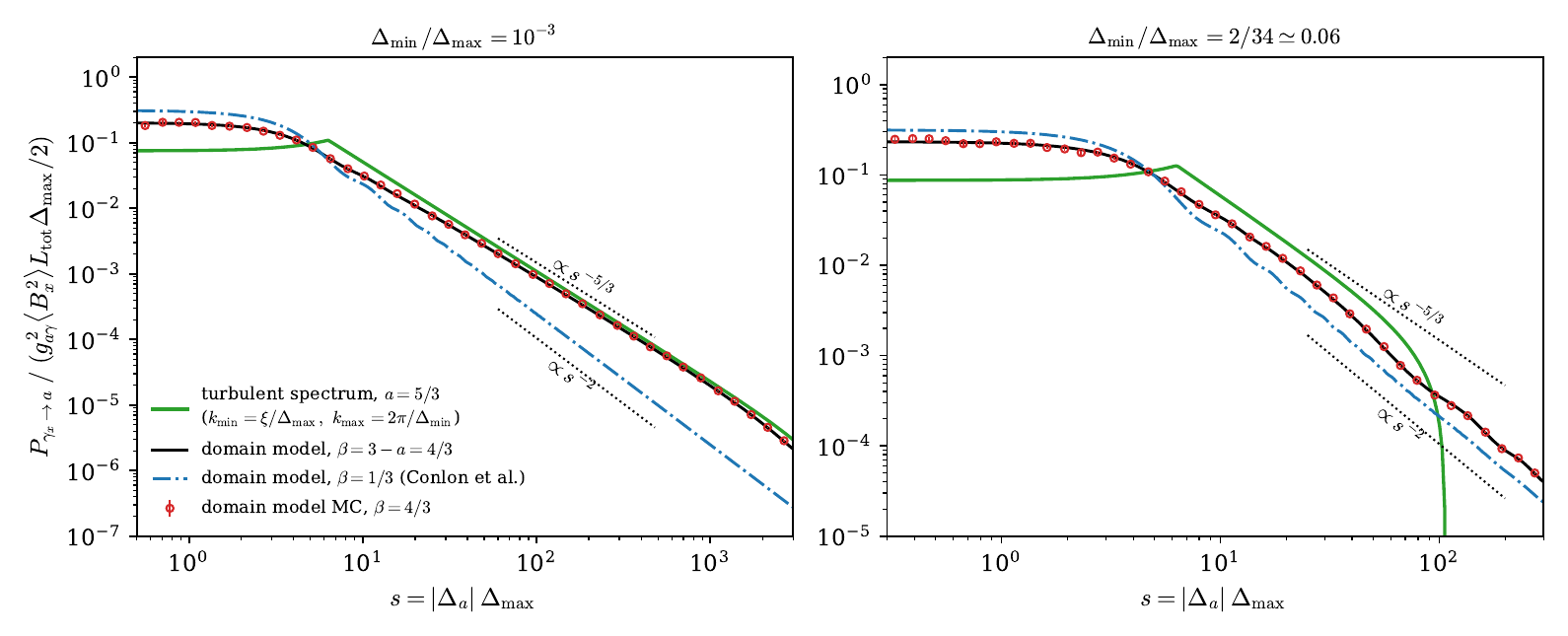}
\caption{Ensemble averaged conversion probability (in units of
$g_{a\gamma}^2\Bx\Ltot\Dmax/2$) as a function of $s=|\Delta_a|\Dmax$, for a
wide range of domain sizes, $\rho=\Dmin/\Dmax=10^{-3}$ (\emph{left}), and for the
realistic value $\rho=2/34\approx0.06$ adopted in~\cite{2025NatAs...9.1387M}
(\emph{right}).
\emph{Green}: turbulent field with Kolmogorov spectrum, $a=5/3$
[\cref{eq:eps_turb}], with $\kmax=2\pi/\Dmin$ and $\kmin=\xi(a)/\Dmax$ from
the amplitude matching \eqref{eq:xi_def}.
\emph{Black}: domain model with $\beta=3-a=4/3$;
\emph{red circles}: direct Monte Carlo simulation of the same domain model
($M=300$ realisations of $\pm B_{x,\rm rms}$ domain fields, log-binned;
error bars show the standard error of the mean).
\emph{Blue dash-dotted}: domain model with $\beta=1/3$, the prescription
of~\cite{Conlon:2017qcw}; its tail falls as $s^{-2}$.
The dotted guides indicate the two slopes.  The spike of the green curve
at $s\approx2\pi$, its lower plateau at small $s$ and, in the right panel,
its drop at $s\approx107$ follow from the sharp cutoffs of \cref{eq:M_def}
(see \cref{sec:turbulent,sec:matching_amplitude}).}
\label{fig:slopes}
\end{figure}

In real space, the power law \eqref{eq:eps_turb_powerlaw} translates into a
non-analytic term in the correlator at small separations.  For
$\kmin z\ll 1 \ll \kmax z$,
\begin{equation}
    1-\frac{K_{\rm turb}(z)}{\Bx}
    = \frac{1}{\pi\Bx}\int_0^\infty dk\,\epsilon_x(k)\left(1-\cos kz\right)
    \simeq
    -\frac{3\,(a^2-1)\,\Gamma(1-a)\sin\frac{\pi a}{2}}{2\,a\,(a+2)}
    (\kmin z)^{a-1},
    \label{eq:structure_turb}
\end{equation}
so the structure function of the field grows as $z^{a-1}$, the standard
scaling of a turbulent cascade (for Kolmogorov, the $z^{2/3}$
law~\cite{Frisch1995}).
The combination $-\Gamma(1-a)\sin(\pi a/2)$ is positive for $1<a<3$.  The
exact correlator for arbitrary $z$, including the oscillatory artefacts
introduced by the sharp cutoffs of \cref{eq:M_def}, is given in closed form in
\cref{app:transforms}.  These artefacts affect the comparison with the domain
model only near $|\Delta_a|\sim\kmin$ and $|\Delta_a|\sim\kmax$
(see \cref{fig:slopes}).

\section{Matching the two descriptions: the \texorpdfstring{$a$--$\beta$}{a-beta} relation}
\label{sec:matching}

In this section we match the line-of-sight spectra of the domain model
(\cref{sec:domain}) and of the turbulent field (\cref{sec:turbulent}).  We
first match the slopes of their power law tails (\cref{sec:matching_slopes})
and then their amplitudes (\cref{sec:matching_amplitude}).  In
\cref{sec:literature} we compare the result with the prescriptions used in
the literature.

\subsection{Matching the spectral slopes}
\label{sec:matching_slopes}

By \cref{eq:master,eq:P_eps0}, the line-of-sight spectrum of the domain
model is $\epsilon_x^{\rm dom}(k)=2\Bx\Dmax f_\beta(k\Dmax)$, so its
large-$s$ behaviour ($s=k\Dmax\gg1$, wavelengths well inside the range of
domain sizes) is given by \cref{eq:f_asym}.
The non-analytic part
$t^{2-\beta}$ of the correlator \eqref{eq:Kdom_eps0} produces the
$s^{-(3-\beta)}$ power law, which is the collective contribution of domains
of all sizes.  The linear part of the correlator produces the $s^{-2}$ term.
Equivalently, the $s^{-2}$ term comes from the kink of $K_{\rm dom}$ at
$z=0$, which the discontinuities of the field at the domain walls produce.
Comparison with the turbulent result \eqref{eq:eps_turb_powerlaw},
$\epsilon_x\propto k^{-a}$, leads to the following conclusions.

\begin{itemize}
\item For $1<\beta<2$ the first term of \cref{eq:f_asym} dominates
($3-\beta<2$), and matching the exponents gives the relation anticipated in
\cref{eq:main_relation_intro},
\begin{equation}
    3-\beta = a
    \qquad\Longleftrightarrow\qquad
    \beta = 3-a ,
    \label{eq:main_relation}
\end{equation}
which covers turbulent spectra with $1<a<2$.  At the lower edge, $a\to1^+$,
the correspondence becomes marginal in both models at once: $\beta\to2$
makes the $\rho\to0$ limit of the domain correlator singular, while $a\to1$
makes the turbulent variance diverge logarithmically in the infrared.
Kolmogorov turbulence, $a=5/3$, corresponds to
\begin{equation}
    \beta = \frac43,
    \qquad
    p(\Delta)\propto\Delta^{-4/3} .
    \label{eq:kolmogorov_beta}
\end{equation}
The same conclusion follows in real space from matching the structure
functions \eqref{eq:structure_dom} and \eqref{eq:structure_turb}, in which
$t^{2-\beta}$ corresponds to $z^{a-1}$.
\item For $\beta<1$ the $s^{-2}$ term dominates at large $s$ for any
$\beta$, and the field discontinuities pin the spectrum of the domain model to
the slope $-2$.  The slope is universal, while the amplitude
$(2-\beta)/(1-\beta)$ of the $s^{-2}$ term still depends on $\beta$ and
approaches the value $1$ of uniform domains as $\beta\to-\infty$.  Steeper
spectra, $2<a<3$, cannot be
represented by this class of models at all.\footnote{This slope-$(-2)$
``floor'' is the spectral counterpart of the observation
of~\cite{Matthews2022} (see also~\cite{Galanti:2022ijh}) that the kinks of
the autocorrelation functions of domain models enhance the conversion
probability at high $|\Delta_a|$, i.e.\ at low energies.}
\item The boundary case $\beta=1$ (log-uniform domain sizes,
\cref{eq:Kdom_beta0}) corresponds to $a=2$ up to logarithmic corrections:
the tail of the transform is $(\ln s+\gamma_E)/s^2$, with $\gamma_E$ the
Euler--Mascheroni constant [see \cref{eq:f_beta0}].
\end{itemize}

The left panel of \cref{fig:slopes} illustrates the first two statements.
Throughout the inertial range the domain model with $\beta=3-a=4/3$
reproduces the slope of the turbulent $a=5/3$ conversion probability and,
once the cutoffs are identified as derived in \cref{sec:matching_amplitude},
also the amplitude of its power law tail.  The model with $\beta=1/3$ instead
falls with the universal slope $-2$ and deviates from the turbulent curve by a
factor that grows towards large $s$ (low energies).  We note that the negative $s^{-2}$ term in
\cref{eq:f_asym} biases single power law fits of the $\beta=4/3$ tail
towards shallower slopes, $\approx-1.62$ instead of $-5/3$ for a fit over
$100\lesssim s\lesssim1000$.  The index extracted from simulations should therefore be
fitted with the two-term form \eqref{eq:f_asym}.  Outside the inertial
range the correspondence degrades, as quantified in
\cref{sec:matching_amplitude}.  The right panel shows the same comparison for
the realistic scale ratio $\rho\approx0.06$, for which the inertial range
spans only about one decade (see \cref{sec:literature}).

\begin{figure}[t]
\centering
\includegraphics[width=0.6\linewidth]{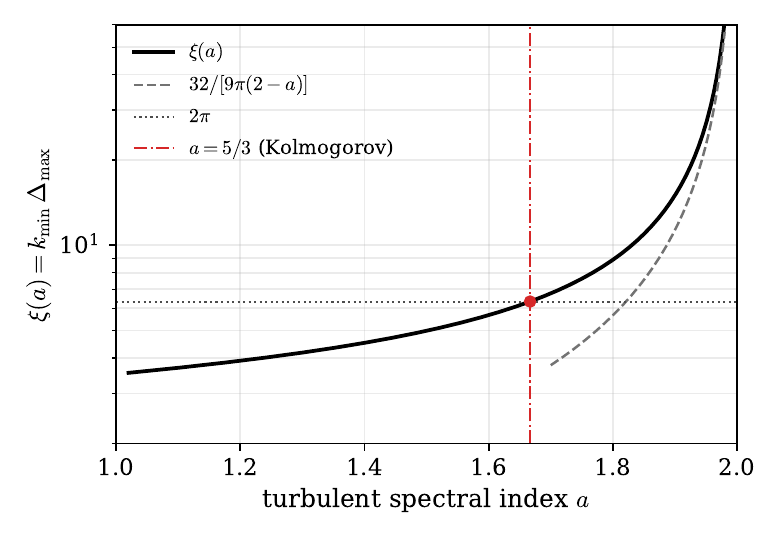}
\caption{Cutoff matching factor $\xi(a)=\kmin\Dmax$ of \cref{eq:xi_def} as a
function of the turbulent spectral index $a$ (\emph{solid line}).  The dotted
line marks $2\pi$; the dash-dotted line and the circle mark the Kolmogorov
value $a=5/3$, where $\xi=6.33$.  The dashed line is the leading behaviour
$\xi\simeq32/[9\pi(2-a)]$ near $a\to2$, where the two power law tails of
\cref{eq:f_asym} merge and the amplitude matching degenerates.}
\label{fig:xi}
\end{figure}
\subsection{Matching the amplitudes: relation between the cutoffs}
\label{sec:matching_amplitude}

With $\beta=3-a$, the power law tails of \cref{eq:f_asym} and
\cref{eq:eps_turb_powerlaw} have the same slope.  Equating also their
amplitudes fixes the combination $\kmin\Dmax$ and determines which spectral
cutoff a given maximal domain size represents.  Writing $\kmin =
\xi(a)/\Dmax$, we find
\begin{equation}
    \xi(a)
    = \left[
      \frac{4\,a\,(a+2)\Gamma(a)\cos\frac{\pi a}{2}}
           {3\pi\,(a-2)\,(a^2-1)}
      \right]^{\tfrac{1}{a-1}} .
    \label{eq:xi_def}
\end{equation}
\Cref{fig:xi} shows the function $\xi(a)$.  For $a\to2^-$ it diverges as
$\xi\simeq32/[9\pi(2-a)]$, since there
the two tails of \cref{eq:f_asym} merge and the amplitude matching
degenerates [see \cref{eq:f_beta0}].

For the Kolmogorov value
$a=5/3$,
\begin{equation}
    \kmin = \frac{6.33}{\Dmax} = \frac{2\pi}{\Dmax}\times1.007 ,
\end{equation}
so the amplitude matching reproduces the naive identification
$\kmin=2\pi/\Dmax$ to better than one per cent.  We are not aware of a deeper
reason for this coincidence; for other values of $a$ the deviation from
$2\pi$ reaches tens of per cent (\cref{fig:xi}).  The corresponding
identification at the small scale end, $\kmax=2\pi/\Dmin$, affects only the
region $|\Delta_a|\gtrsim\kmax$.  There the sharply cut turbulent spectrum
ends, while the domain model keeps the $k^{-2}$ tail produced by its walls.
This region lies at the lowest energies, far outside the inertial range, so
the precise choice of $\kmax$ does not affect our conclusions.

\paragraph{The saturation plateau.}
At $s\lesssim\xi$ (high photon energies) both conversion
probabilities saturate, but at different levels (\cref{fig:slopes}).  The plateau of the domain model is
$f_\beta(0)=(2-\beta)/[2(3-\beta)]=(a-1)/(2a)$ in units of $2\Bx\Dmax$
[\cref{eq:f_zero}], while for the sharply cut spectrum
$\epsilon_x(0)/(2\Bx\Dmax)=3\pi(a-1)/(8a\xi)$; their ratio is
\begin{equation}
    R(a) = \frac{4\,\xi(a)}{3\pi}
    = \{1.7,\ 1.9,\ 2.7,\ 3.8\}
    \quad\text{for}\quad a=\{1.2,1.4,5/3,1.8\} .
    \label{eq:plateau_ratio}
\end{equation}
The mismatch arises because the smallest wavenumbers, $k\sim\kmin$,
dominate the plateau $\epsilon_x(0)=2\pi^2\int k\,\mathcal{M}(k)\,dk$.  These are scales
at and beyond the outer scale, where neither the sharp spectral cutoff nor
the maximal domain size describes a real cluster field faithfully.
For example, an exponential infrared taper,
$\mathcal{M}(k)\propto k^{-(a+2)}e^{-\kmin/k}$, normalised to the same $\Bx$,
in place of the sharp cutoff raises the turbulent plateau by exactly a factor $a$ ($5/3$ for Kolmogorov
turbulence) and reduces $R$ from $2.7$ to $1.6$.  The inertial range
amplitude then shifts by $1/\Gamma(a)\approx1.1$, an ambiguity of the same
infrared origin.  The plateau height ($p_0$ in the notation
of~\cite{2025NatAs...9.1387M}) is therefore uncertain by a factor of order
unity, and in applications it should be calibrated against the measured
field variance and correlation length [\cref{eq:Lc}].

\subsection{Comparison with the literature}
\label{sec:literature}

\paragraph{The $\beta=1/3$ prescription.}
Ref.~\cite{Conlon:2017qcw} draws domain lengths from
$p(\Delta)\propto\Delta^{-1/3}$, stating that ``this index
is based on the Kolmogorov power spectrum ${\rm Prob}(|k|)\propto k^{-5/3}$''.
This value follows if one treats the energy spectrum as a probability
distribution over scales and applies the change of variables
$\Delta\sim1/k$, $E(k)\,dk\propto k^{-5/3}dk \mapsto
\Delta^{5/3}\,\Delta^{-2}\,d\Delta = \Delta^{-1/3}d\Delta$.  The
statistics that the domain sequence generates along the sightline follow
from its correlator \eqref{eq:Kdom}.  For $\beta=1/3$ the field has a
structure function that grows linearly at small separations (the first term
of \cref{eq:structure_dom} dominates) and a line-of-sight spectrum falling as
$k^{-2}$, both signatures of an $a=2$ spectrum (\cref{sec:matching_slopes}).
Requiring the two-point statistics of the generated field to reproduce those
of a Kolmogorov cascade gives $\beta=3-a=4/3$, one unit away in the
exponent.

\paragraph{The \textsc{ALPro} default.}
Ref.~\cite{2020ApJ...890...59R} and the public code
\textsc{ALPro}~\cite{alpro,Matthews2022} sample domain sizes from
$p(\Delta)\propto\Delta^{-1.2}$; these references do not
derive this choice.  By \cref{eq:main_relation}, $\beta=1.2$ corresponds to
the spectral index $a=1.8$ ($n=a+2=3.8$).  This value is close to the
Kolmogorov value $a=5/3$ ($n=11/3$), which corresponds to
$p(\Delta)\propto\Delta^{-4/3}$.

\paragraph{Which searches are sensitive to the index?}
How much the choice of $\beta$ matters depends on the range of the variable
$s=|\Delta_a|\Dmax$ [\cref{eq:s_def}] that a given analysis probes.  For the
gamma ray ensemble analysis of~\cite{2025NatAs...9.1387M}
($m_a\approx3\,$neV, $\Dmax=34\,$kpc, and the range $\omega=1$--$100\,$GeV
used in that analysis) the variable spans $s\approx0.24$--$24$.  This range
covers both the plateau and a substantial part of the inertial range, so both
the index and the amplitude ambiguity of \cref{sec:matching_amplitude} are
directly relevant.
By contrast, X-ray point source
searches~\cite{Conlon:2017qcw,2020ApJ...890...59R,2022MNRAS.510.1264S}
target very light ALPs, $m_a\lesssim10^{-12}\,$eV.  For such masses
$m_a$ lies below the plasma frequency everywhere in the cluster, and the
momentum mismatch is set by the plasma term (cf.\ the footnote in
\cref{sec:master}).  The relevant variable is then
\begin{equation}
    s\simeq\frac{\omega_{\rm pl}^2\Dmax}{2\omega} .
    \label{eq:s_plasma}
\end{equation}
For $\omega=3\,$keV and $\Dmax=34\,$kpc it is of order unity at
$n_e=10^{-3}\,{\rm cm^{-3}}$ and can grow to
$s\sim10$--$10^{2}$ in the dense regions of the cluster
(see~\cite{Conlon:2017qcw,2020ApJ...890...59R}).  Such searches therefore
probe both the plateau and the inertial range of the averaged conversion
probability.  Since $\omega_{\rm pl}$ follows the density profile,
$\Delta_a=(\omega_{\rm pl}^2-m_a^2)/(2\omega)$ varies along the line of
sight, and our formulae for constant $\Delta_a$ apply only locally.

\paragraph{Implications for the ensemble analysis.}
The ensemble averaged analysis of~\cite{2025NatAs...9.1387M} explored
domain size distributions from uniform domains ($\beta=0$) to the
$\beta=1/3$ distribution of~\cite{Conlon:2017qcw}.  This whole range belongs
to the $\beta<1$ class of \cref{sec:matching_slopes}.  At the scale ratio
$\rho=2/34$ of that analysis, the averaged conversion probability for the
derived index $\beta=4/3$ differs from the $\beta=1/3$ case by a factor of
$0.74$ at the high energy end of the band ($s\lesssim1$), crosses unity at
$s\approx5$, and reaches $1.9$ at the low energy end ($s\approx24$), see the
right panel of \cref{fig:slopes}.  For a cluster with a measured spectral
index $a$ (equivalently $n=a+2$), the relation \eqref{eq:main_relation} fixes
the domain model index to $\beta=3-a$, and the remaining modelling freedom
lies in the outer scale amplitude discussed in
\cref{sec:matching_amplitude}.

\section{Physical interpretation of the relation
\texorpdfstring{$\beta=3-a$}{beta = 3-a}}
\label{sec:interpretation}

In this section we give a physical interpretation of the relation
\eqref{eq:main_relation}, which at first sight seems counterintuitive.  A
turbulent spectrum with $1<a<2$ places most of the magnetic energy at the
largest scales, since the energy per logarithmic interval of wavenumber,
$k\,E_B(k)\propto k^{1-a}$, grows towards small $k$.  According to
\cref{eq:main_relation}, such a field is represented by the domain size
distribution $p(\Delta)\propto\Delta^{-(3-a)}$, which is dominated by the
smallest domains.  Indeed, the number of domains per logarithmic interval of
size scales as $\Delta\,p(\Delta)\propto\Delta^{a-2}$.  For Kolmogorov
turbulence every decade down in scale therefore contains
$10^{2-a}\approx2.2$ times more domains, and the number density $p(\Delta)$
itself grows by a factor $10^{3-a}\approx22$.  Below we explain how a size
distribution dominated by small domains can represent a field whose power
resides at large scales.

\paragraph{Number versus path length.}
A domain of size $\Delta$ occupies a stretch $\Delta$ of the sightline, so
the two-point function weights each scale by the path length that domains of
this scale cover.  The fraction of the sightline residing in
domains of scale $\sim\Delta$, per logarithmic interval, is
\begin{equation}
    \frac{d\mathcal{F}}{d\ln\Delta}
    = \frac{\Delta^{2}\,p(\Delta)}{\avg{\Delta}}
    \propto \Delta^{2-\beta} = \Delta^{a-1},
    \label{eq:coverage}
\end{equation}
which increases towards large scales.  For the Kolmogorov value
$\beta=4/3$ and $\rho\to0$, $79\%$ of the path lies inside domains within
one decade of $\Dmax$, and half of the path lies inside domains larger than
$2^{-3/2}\Dmax\approx0.35\,\Dmax$.  The correlation length is
correspondingly large, $L_c=0.4\,\Dmax$ [\cref{eq:Lc}].  A similar
situation occurs in the cascade itself, where small eddies outnumber the
large ones but carry little of the energy~\cite{Frisch1995}.

\paragraph{Energy per logarithmic interval.}
Because every domain carries the same field variance $\avg{B_x^2}$, the
magnetic energy per logarithmic interval of scale is proportional to the
path length that domains of this scale cover, \cref{eq:coverage}.  The turbulent
cascade distributes its energy over scales $\lambda\sim1/k$ as
$k\,E_B(k)\propto\lambda^{a-1}$.  Requiring that the two distributions
agree, we obtain
\begin{equation}
    \underbrace{\Delta^{2-\beta}}_{\text{path fraction}}
    =
    \underbrace{\Delta^{a-1}}_{\text{energy fraction}}
    \qquad\Longrightarrow\qquad
    \beta = 3-a ,
    \label{eq:octave_matching}
\end{equation}
which reproduces \cref{eq:main_relation}.  We refer to this reasoning as
the coverage argument.

\paragraph{Amplitude and filling fraction.}
The same argument shows why the domain model needs the large number of small
domains.  In a real cascade, structures of scale $\ell$ are present
everywhere but are weak: the field difference across a separation $\ell$
scales as $\delta B(\ell)\propto\ell^{(a-1)/2}$, which is $\ell^{1/3}$ for
Kolmogorov turbulence.  The domain model cannot vary its amplitude with
scale, since every domain carries the full field strength.  It can represent
weak structure at small scales only by making such structure rare, with
domains of scale $\ell$ and full field strength covering a fraction
$\propto\ell^{a-1}$ of the path.  The two-point statistics do not distinguish
weak fluctuations present everywhere from strong fluctuations that cover a
small fraction of the path.  Since each domain of scale $\ell$ covers only a
length $\ell$, the coverage $\propto\ell^{a-1}$ requires the number of
domains per logarithmic interval to scale as $\ell^{a-1}/\ell=\ell^{a-2}$, which diverges
towards small scales for $a<2$.  This construction resembles the
intermittency models of turbulence, in which eddies of scale $\ell$ fill a
decreasing fraction of the volume~\cite{Frisch1995}.

\paragraph{Validity of the coverage argument.}
The coverage argument \eqref{eq:octave_matching} assumes \emph{scale
locality}: a domain of size $\Delta$ contributes spectral power only at
$k\sim1/\Delta$.  A domain is a
top-hat, and its sharp walls produce the universal $k^{-2}$ tail at all
higher wavenumbers.  The line-of-sight spectrum at a given $k$ therefore
receives two contributions [\cref{eq:f_asym}].  The domains of size
$\sim1/k$ contribute $\propto k^{-(3-\beta)}$, and the walls of all larger
domains contribute $\propto k^{-2}$.  For $1<\beta<2$ the first contribution
dominates, and the coverage argument is justified a posteriori.  For $\beta<1$
the walls dominate and the argument fails: it would predict the effective
index $a_{\rm eff}=3-\beta>2$, whereas the true spectrum has
$a_{\rm eff}=2$ (\cref{sec:matching_slopes}).  The same two contributions
appear in real space, where the structure function of a domain field
measures the probability that a separation $z$ straddles a domain wall.
When the mean domain size is finite ($\beta<1$), the walls form a
quasi-Poisson set, and this probability grows linearly in $z$, as for the
$a=2$ correlator.  The sub-linear growth $z^{a-1}$ of the Kolmogorov
structure function requires the scale-free clustering of walls that the
distributions with $1<\beta<2$ provide.  Since the probability to cross a wall
cannot grow more slowly than linearly at small $z$, spectra steeper than
$a=2$ are out of reach.  This is the real space counterpart of the $k^{-2}$
floor of \cref{sec:matching_slopes}.

We believe that this also explains how heuristics that assign domain
numbers by energy content can lead to different prescriptions.  The change of variables that gives $\beta=1/3$
(\cref{sec:literature}) assigns the number of domains per logarithmic interval according to
the energy content at each scale.  The length bias of \cref{eq:Pi_renewal}
then weights the large scales a second time, and the walls remain the only
small scale structure of the resulting field.

\section{Energy dependence of the averaged signal}
\label{sec:analytic}

In this section we relate the averaged conversion probability of
\cref{sec:analytic_dom} to the step-like feature of the ensemble analyses.
Restoring physical units via \cref{eq:s_def}, we write it as a function of
photon energy $\omega$,
\begin{equation}
    \avg{P_{\gamma\to a}}(\omega)
    = \frac{g_{a\gamma}^2\,\Bx\,\Ltot\,\Dmax}{2}
      f_\beta\left(\frac{m_a^2\Dmax}{2\omega}\right) .
    \label{eq:P_energy}
\end{equation}
\Cref{eq:P_energy} describes analytically the smooth, step-like absorption
feature identified in~\cite{2025NatAs...9.1387M}.  The suppression of the
stacked spectrum switches on around the critical energy
$\omega_c \sim m_a^2\Dmax/2$, where $s\sim1$.  Towards higher energies it
grows as the power law $\omega^{\,3-\beta}=\omega^{\,a}$ [see
\cref{eq:f_asym,eq:main_relation}], or as $\omega^2$ for models with $\beta<1$, and it
saturates at $\avg{P}=g_{a\gamma}^2\Bx\Ltot L_c/4$ for $\omega\gg\omega_c$.
The sigmoid-type fitting formulae of~\cite{2025NatAs...9.1387M} describe
this feature empirically, while \cref{eq:P_energy} expresses its onset energy
and power law rise through $m_a$, $\Dmax$ and $a$.

\paragraph{From idealised to realistic clusters.}
In real galaxy clusters the strength of the magnetic field typically
decreases gradually from the centre of the cluster towards its
outskirts~\cite[see e.g.][]{Bonafede:2010xg,2018MNRAS.481.5046R},
and the domain lengths may correlate with the distance from the cluster
centre.  We expect such spatial dependencies to modify the relation
$\beta=3-a$ and the conversion probability \eqref{eq:P_energy} to some
extent.  The field at the largest scales $\Dmax$ is, however, expected to
vary only slowly, which suggests that these modifications are weak.  We leave
a detailed study of realistic magnetic field profiles to future work.

\section{Conclusions}
\label{sec:conclusions}

Averaging the photon--ALP conversion over many sightlines has been
used for ALP searches in galaxy clusters~\cite{2025NatAs...9.1387M,2025arXiv250922344M}.
The average replaces the stochastic spectral modulations of individual
sightlines by a smooth suppression feature.  In the perturbative
regime the averaged signal is fully determined by the line-of-sight
power spectrum of the transverse magnetic field~\cite{Marsh:2021ajy}.  In this work we used this property to answer
analytically a question relevant for all simulations that model turbulent
cluster fields by domains: which domain size distribution corresponds
to a given turbulent spectrum.

For a statistically homogeneous field, the domain model with
$p(\Delta)\propto\Delta^{-\beta}$ reproduces the averaged conversion
probability of a turbulent field with $\avg{|B_k|^2}\propto k^{-(a+2)}$ if
$\beta=3-a$, for $1<a<2$.  Kolmogorov turbulence therefore corresponds to
$\beta=4/3$ (cf.\ the value $\beta=1/3$ prescribed in~\cite{Conlon:2017qcw}
and adopted in subsequent
analyses~\cite{2025NatAs...9.1387M,2026JCAP...07..066G}).  Physically, the relation
states that the fraction of the sightline covered by domains of a given
scale equals the fraction of the turbulent energy per logarithmic interval
of scale; both scale as $\Delta^{a-1}$.  The numerous small domains represent
the weak small scale fluctuations of the cascade by rare structures of full
field strength.  This interpretation holds for $1<a<2$, where the collective
contribution of the domains dominates over the $k^{-2}$ contribution of the
domain walls (\cref{sec:interpretation}).

Domain models with $\beta<1$ mimic an $a=2$ spectrum for any $\beta$,
because the field discontinuities at the domain walls produce a $k^{-2}$
tail of the line-of-sight power spectrum.  Spectra steeper than $a=2$ cannot
be represented by domain models at all.  Matching the amplitudes of the power
law tails further relates the spectral cutoff to the largest domain size,
$\kmin=\xi(a)/\Dmax$ with $\xi$ given by \cref{eq:xi_def}.  For Kolmogorov
turbulence $\xi$ agrees with $2\pi$ to better than one per cent.  Finally,
we obtained the ensemble averaged conversion probability of the domain model
in closed form, \cref{eq:f_closed}, with the case of arbitrary $\Dmin/\Dmax$
given in \cref{eq:I1_app,eq:I2_app}, and derived the two power law
approximation \eqref{eq:f_asym}.  Together they describe analytically the
step-like feature used in~\cite{2025NatAs...9.1387M}: its onset at
$\omega_c\sim m_a^2\Dmax/2$, its power law rise $\propto\omega^{a}$, and its
saturation, which the correlation length \eqref{eq:Lc} controls.

The relation $\beta=3-a$ fixes only the spectral shape of the ensemble
averaged signal, which the inertial range controls.  The
absolute level of the saturation at high energies is a property of the
outer scale and carries a modelling uncertainty of order unity for any
cutoff prescription [\cref{eq:plateau_ratio} and the discussion below it].
Our formulae hold in the perturbative regime (\cref{app:qm}) and for a
statistically homogeneous field (\cref{sec:master}).  Within this scope,
the measured turbulent spectral index fixes the domain size index, which the
propagation codes treat as an \emph{ad hoc} parameter.  Our
explicit formulae also replace the Monte Carlo averaging over field
realisations.  We expect them to be useful both for reinterpreting the ensemble analysis
of~\cite{2025NatAs...9.1387M} with the derived index $\beta=4/3$ and for
analysing future ensemble searches.

The scatter of the conversion probability between sightlines, which
involves four-point functions of the field (\cref{sec:master}), the effect of
the derived index on the limits published in~\cite{2025NatAs...9.1387M}, and
the radial profiles of the cluster field and electron density
(\cref{sec:analytic}) lie outside the scope of this work, and we leave them
for future work.

\acknowledgments

The authors thank Denys Tvardovskyi for helpful discussions and for assistance with exploratory calculations. The work of O. Khasai was supported by the Ministry of Education and Science of Ukraine under the project ``Development of Theoretical and Computational Methods for the Study of Micro- and Macrosystems in Quantum Technologies''. O. Khasai gratefully acknowledges the hospitality of the Niels Bohr
Institute, University of Copenhagen, where a significant part of this
research was carried out during a traineeship supported by the Erasmus+
programme of the European Union, and warmly thanks Stanislav Vilchinskii,
Danylo Komisar and Yelyzaveta Komisar for their support.
The authors acknowledge support by the state of Baden-W\"urttemberg through bwHPC.
This project has received funding through the MSCA4Ukraine project, grant number Ref 1.4 - UKR - 1245772 - MSCA4Ukraine, which is funded by the European Union. Views and opinions expressed are however those of the author(s) only and do not necessarily reflect those of the European Union, the European Research Executive Agency or the MSCA4Ukraine Consortium. Neither the European Union nor the European Research Executive Agency, nor the MSCA4Ukraine Consortium as a whole nor any individual member institutions of the MSCA4Ukraine Consortium can be held responsible for them. The work of L. Zadorozhna was partially supported by the Ministry of Education and Science of Ukraine under the project ``Search for dark matter and particles beyond the Standard Model'' (0125U002260). This research was conducted with support from the Centre for Collective Use of Scientific Equipment ``Laboratory of High Energy Physics and Astrophysics'', Taras Shevchenko National University of Kyiv.

\newpage
\appendix
\section{Photon--ALP mixing and the master formula}
\label{app:qm}

\subsection{Equations of motion and perturbative solution}

In the relativistic, quasi-classical limit, propagation of the photon--ALP
system along $z$ is governed by a Schr\"odinger-like
equation~\cite{1988PhRvD..37.1237R},
\begin{equation}
    i\frac{d\Psi}{dz} = (H_0+H_I)\,\Psi,
    \qquad
    \Psi = \begin{pmatrix} A_x\\ A_y\\ \phi\end{pmatrix},
\end{equation}
with
\begin{equation}
    H_0 = {\rm diag}\left(\omega,\omega,\omega+\Delta_a\right),
    \qquad
    H_I = \begin{pmatrix} 0&0&\Delta_x\\ 0&0&\Delta_y\\
    \Delta_x&\Delta_y&0\end{pmatrix},
\end{equation}
where $A_{x,y}$ are the photon polarisation amplitudes, $\phi$ the ALP
amplitude, and
\begin{equation}
    \Delta_a = -\frac{m_a^2}{2\omega},
    \qquad
    \Delta_{x,y}(z) = \frac{g_{a\gamma} B_{x,y}(z)}{2}.
\end{equation}
We pass to the interaction picture,
$\Psi_{\rm int}=e^{izH_0}\Psi$, and expand the evolution operator to first
order in $H_I$.  The $\gamma_x\to a$ amplitude after a path of length $\Ltot$
is then
\begin{equation}
    \mathcal{A}_{\gamma_x\to a}
    = -i\int_0^{\Ltot} dz\,\Delta_x(z)\,e^{i\Delta_a z} ,
    \label{eq:amplitude_app}
\end{equation}
whose squared modulus gives \cref{eq:P_perturbative}.

\subsection{Ensemble average}

Squaring \cref{eq:amplitude_app} and averaging over field realisations,
\begin{equation}
    \avg{P_{\gamma_x\to a}}
    = \frac{g_{a\gamma}^2}{4}
    \int_0^{\Ltot}dz_1\int_0^{\Ltot}dz_2\,
    \avg{B_x(z_1)B_x(z_2)}\, e^{i\Delta_a(z_1-z_2)} .
\end{equation}
For a statistically homogeneous field,
$\avg{B_x(z_1)B_x(z_2)}=K(|z_1-z_2|)$, and changing variables to
$u=z_1-z_2$,
\begin{equation}
    \avg{P_{\gamma_x\to a}}
    = \frac{g_{a\gamma}^2}{4}
    \int_{-\Ltot}^{\Ltot} du\,\left(\Ltot-|u|\right) K(|u|)\,
    e^{i\Delta_a u}
    \xrightarrow[\Ltot\gg L_c]{}
    \frac{g_{a\gamma}^2\,\Ltot}{2}
    \int_0^{\infty} du\,K(u)\cos(\Delta_a u),
\end{equation}
which is \cref{eq:master}; the finite path correction is suppressed by
$L_c/\Ltot$.  For the domain model the equivalence between the average over
configurations and the average over the distributions of domain sizes and
field values follows from the law of large numbers, and is verified
numerically in \cref{fig:average}.

\subsection{Validity of perturbation theory}
\label{app:validity}

For a stochastic field the natural expansion parameter of the
perturbation series is the accumulated conversion probability itself.  The
first order amplitude \eqref{eq:amplitude_app} is a sum of
$\sim\Ltot/L_c$ quasi-independent contributions, and its ensemble variance
is, by construction, exactly the averaged probability \eqref{eq:master}.
Neglecting the depletion of the photon beam (the secular, second order
$\gamma\to a\to\gamma$ terms) is therefore justified when
\begin{equation}
    \avg{P_{\gamma\to a}}\ll1 ,
    \qquad\text{i.e.\ at the plateau [\cref{eq:f_zero,eq:Lc}]}\quad
    \frac{g_{a\gamma}^2\,\Bx\,\Ltot\,L_c}{4}\ll1 .
    \label{eq:pt_cond2}
\end{equation}
Numerically,
\begin{equation}
    \avg{P}(0)
    \approx 0.023
    \left(\frac{g_{a\gamma}}{10^{-12}\,{\rm GeV^{-1}}}\right)^{2}
    \left(\frac{B_{x,\rm rms}}{1\,\mu{\rm G}}\right)^{2}
    \frac{\Ltot}{1\,{\rm Mpc}}
    \frac{L_c}{10\,{\rm kpc}} ,
    \label{eq:P_numeric}
\end{equation}
so the couplings probed by current cluster searches
($g_{a\gamma}\lesssim\text{few}\times10^{-12}\,{\rm GeV^{-1}}$) are
perturbative for typical field strengths, while the largest values of the
field and the coupling approach saturation.  In addition, the mixing
must be weak over each correlation length: the amplitude accumulated over
one correlation length is
$\sim\Delta_{a\gamma}\min(L_c,|\Delta_a|^{-1})$ with
$\Delta_{a\gamma}=g_{a\gamma}B_\perp/2$, giving
\begin{equation}
    \Delta_{a\gamma}\ll\max\left(|\Delta_a|,L_c^{-1}\right) .
    \label{eq:pt_cond1}
\end{equation}
In the oscillatory regime, $|\Delta_a|L_c\gg1$, this is the familiar
condition for a homogeneous field, $g_{a\gamma}B_\perp\omega/m_a^2\ll1$.  At
high energies, $|\Delta_a|L_c\ll1$, it becomes $\Delta_{a\gamma}L_c\ll1$.

When \cref{eq:pt_cond2} is violated, depletion must be resummed.  For
domains of equal size the exact result for many domains is
known~\cite{2002PhRvL..88p1302C,2002PhLB..543...23G,2008LNP...741..115M}:
\begin{equation}
    P_{\gamma\to a}=\frac13\left[1-e^{-3NP_0}\right],
    \qquad N=\frac{\Ltot}{L},
    \label{eq:resummed}
\end{equation}
where $N$ is the number of domains along the path and $P_0$ is the single
domain conversion probability of \cref{eq:P0_check}; the factor $3$ in the
exponent includes $\avg{B_\perp^2}=2\Bx$ for randomly oriented domains.
The probability saturates at the equipartition value
$\avg{P}\to1/3$ among the two photon polarisations and the ALP, and its
leading term $NP_0$ is the averaged probability \eqref{eq:P0_check}.

\section{Explicit transforms}
\label{app:transforms}

\subsection{Fourier cosine transform of the domain correlator at arbitrary
\texorpdfstring{$\rho$}{rho}}

For the integral entering the master formula \eqref{eq:master} we write
\begin{equation}
    \int_0^\infty K_{\rm dom}(z)\cos(\Delta_a z)\,dz
    = \Bx\,(I_1+I_2),
    \qquad\text{so that}\qquad
    \avg{P_{\gamma_x\to a}}
    = \frac{g_{a\gamma}^2}{2}\,\Ltot\,\Bx\,(I_1+I_2),
    \label{eq:Fdom_app}
\end{equation}
with $K_{\rm dom}$ from \cref{eq:Kdom} and $q\equiv|\Delta_a|$ below.  The
short separation piece is elementary:
\begin{equation}
    I_1 = \int_0^{\Dmin}\left(1-\frac{z}{\avg{\Delta}}\right)\cos(qz)\,dz
    = \frac{1-\cos(q\Dmin)
    +\left(\avg{\Delta}-\Dmin\right)q\sin(q\Dmin)}{q^2\avg{\Delta}} .
    \label{eq:I1_app}
\end{equation}
For the intermediate piece, using
$1/\avg{\Delta}=(2-\beta)D_0/[(1-\beta) D_1]$ with
$D_0=\Dmax^{1-\beta}-\Dmin^{1-\beta}$, $D_1=\Dmax^{2-\beta}-\Dmin^{2-\beta}$, the
integrand simplifies to
\begin{equation}
    K_{\rm dom}(z)\big|_{\Dmin\le z<\Dmax}
    = \frac{\Bx}{D_1}\left[
    \Dmax^{2-\beta}-\frac{2-\beta}{1-\beta}\Dmax^{1-\beta}\,z
    +\frac{z^{2-\beta}}{1-\beta}\right],
\end{equation}
and therefore
\begin{align}
    I_2 ={}& \frac{1}{D_1}\Bigg[
    \Dmax^{2-\beta}\,\frac{\sin(qz)}{q}
    -\frac{2-\beta}{1-\beta}\,\Dmax^{1-\beta}
    \left(\frac{\cos(qz)}{q^{2}}+\frac{z\sin(qz)}{q}\right)
    \Bigg]_{z=\Dmin}^{z=\Dmax}
    \nonumber\\
    &+\frac{1}{(1-\beta) D_1}\,
    {\rm Re}\left[(-iq)^{-(3-\beta)}
    \Bigl(\Gamma\bigl(3-\beta,-iq\Dmin\bigr)
    -\Gamma\bigl(3-\beta,-iq\Dmax\bigr)\Bigr)\right],
    \label{eq:I2_app}
\end{align}
where we used
$\int_{z_1}^{z_2}z^{2-\beta}e^{iqz}dz
=(-iq)^{-(3-\beta)}[\Gamma(3-\beta,-iqz_1)-\Gamma(3-\beta,-iqz_2)]$.
In the limit $\rho\to0$ (for $\beta<2$) the sum $I_1+I_2$ reduces to
$\Dmax f_\beta(s)$ with $s=q\Dmax$ and $f_\beta$ given by
\cref{eq:f_closed}, and \cref{eq:Fdom_app} reduces to \cref{eq:P_eps0}.

\subsection{Line-of-sight spectrum and correlator of the turbulent field}

Setting $\bm{r}=z\hat{\bm z}$ in \cref{eq:M_def} gives
$K_{\rm turb}(z)=\int d^3k\,(1-\hat k_x^2)\mathcal{M}(k)\,e^{ik_zz}$; inserting this
into the definition \eqref{eq:eps_def} produces $\int dz\,
e^{i(k_z'-k_z)z}=2\pi\delta(k_z'-k_z)$, so the line-of-sight spectrum of
$B_x$ is the transverse integral
\begin{equation}
    \epsilon_x(k_z)
    = 2\pi\int dk_x\,dk_y \left(1-\hat k_x^2\right) \mathcal{M}(k)
    = (2\pi)^2\int_0^\infty k_\perp\,dk_\perp
    \left(1-\frac{k_\perp^2}{2k^2}\right)\mathcal{M}(k) ,
\end{equation}
where the azimuthal average replaced $\hat k_x^2\to k_\perp^2/(2k^2)$ and
$k^2=k_\perp^2+k_z^2$.  Changing the variable to $k$ ($k\,dk=k_\perp
dk_\perp$ at fixed $k_z$) yields \cref{eq:eps_turb_int} and, upon integration,
\cref{eq:eps_turb}.  The averaged conversion probability of the turbulent
model plotted in \cref{fig:slopes} is
$\avg{P}=g_{a\gamma}^2\Ltot\,\epsilon_x(\Delta_a)/4$.

The real space correlator is the inverse transform
$K_{\rm turb}(z)=\frac{1}{\pi}\int_0^\infty\epsilon_x(k)\cos(kz)\,dk$;
performing instead the angular integration of the three-dimensional
transform directly gives the equivalent closed form
\begin{equation}
    K_{\rm turb}(z)
    = \frac{3\,\Bx\,(a-1)}{2\left(\kmin^{1-a}-\kmax^{1-a}\right) z^{1-a}}
    \left[
    \frac{\sin x - x\cos x}{(a+2)\,x^{a+2}}
    + \frac{a+1}{a+2}\,
    {\rm Im}\left(i^{\,a}\,\Gamma(-a,\,ix)\right)
    \right]_{x=z\kmin}^{x=z\kmax} .
    \label{eq:Kturb_closed}
\end{equation}
Expanding \cref{eq:Kturb_closed} (or the defining integral) for
$\kmin z\ll1\ll\kmax z$ produces the structure function scaling
\cref{eq:structure_turb}.  The sharp cutoffs of \cref{eq:M_def} generate
oscillatory artefacts $\propto\cos(\kmax z)$, $\cos(\kmin z)$ in
\cref{eq:Kturb_closed} which have no counterpart in a physical cascade; in
the conversion probability they appear only near $|\Delta_a|\simeq\kmin$
(the spike in \cref{fig:slopes}) and $|\Delta_a|\simeq\kmax$.

\subsection{Notation summary}

\begin{center}
\begin{tabular}{ll}
\toprule
$\omega$, $m_a$, $g_{a\gamma}$ & photon energy, ALP mass, ALP--photon coupling\\
$\Delta_a=-m_a^2/2\omega$ & ALP--photon momentum mismatch (in vacuum)\\
$s=|\Delta_a|\Dmax$ & dimensionless inverse energy, \cref{eq:s_def}\\
$K(z)$, $\epsilon_x(k)$ & line-of-sight correlator and power spectrum,
\cref{eq:K_def,eq:eps_def}\\
$p(\Delta)\propto\Delta^{-\beta}$ & domain size distribution on
$[\Dmin,\Dmax]$\\
$\rho=\Dmin/\Dmax$ & ratio of the smallest to the largest domain size\\
$\mathcal{M}(k)\propto k^{-(a+2)}$ & 3D spectrum on $[\kmin,\kmax]$;
energy spectrum $\propto k^{-a}$; $n=a+2$\\
$\avg{\Delta}$, $L_c$ & mean domain length \eqref{eq:mean_delta},
correlation length \eqref{eq:Lc}\\
$f_\beta(s)$ & dimensionless averaged conversion probability,
\cref{eq:f_closed}\\
$\xi(a)=\kmin\Dmax$ & cutoff matching, \cref{eq:xi_def}\\
\bottomrule
\end{tabular}
\end{center}

\bibliographystyle{JHEP}
\bibliography{sn-bibliography}

\end{document}